\documentclass[11pt]{article}
\usepackage{amsfonts}
\usepackage{amsmath}
\usepackage{amssymb}
\usepackage{indentfirst}
\usepackage{graphicx}
\begin{document}

\title{New family of Maxwell like algebras}
\author{P. K. Concha$^{1,2}$\thanks{patillusion@gmail.com},\, R. Durka$^{3}$\thanks{remigiuszdurka@gmail.com},\, N. Merino$^{3}$\thanks{nemerino@gmail.com},\, E. K. Rodr\'{\i}guez$^{1,2}$\thanks{everodriguezd@gmail.com}, \\
{\small $^{1}$\textit{Departamento de Ciencias, Facultad de Artes Liberales y
}}\\{\small \textit{Facultad de Ingenier\'{\i}a y Ciencias, Universidad Adolfo
Ib\'{a}\~{n}ez,}}\\{\small Av. Padre Hurtado 750, Vi\~{n}a del Mar, Chile}\\{\small $^{2}$\textit{Instituto de Ciencias F\'{\i}sicas y Matem\'{a}ticas,
Universidad Austral de Chile,}}\\{\small Casilla 567, Valdivia, Chile}\\{\small $^{3}$\textit{Instituto de F\'{\i}sica, Pontificia Universidad
Cat\'{o}lica de Valpara\'{\i}so,} }\\{\small Casilla 4059, Valpara\'{\i}so, Chile}}
\maketitle

\vspace{-8cm}
\begin{flushright}
{\footnotesize UAI-PHY-16/01}
\end{flushright}
\vspace{7cm}
\maketitle

\begin{abstract}
We introduce an alternative way of closing Maxwell like algebras. We show,
through a suitable change of basis, that resulting algebras are given by the direct sums of the AdS and the Maxwell algebras already known in the
literature. Casting the result into the $S$-expansion method framework
ensures the straightaway construction of the gravity theories based on a
found enlargement.
\end{abstract}

\section{Introduction}

While contractions and (corresponding to an inverse process) deformations
share the property of preserving the dimension of the Lie algebra, there are
some procedures that allow us to find algebras with a greater number of
generators, even in a way that the original algebra is not necessarily
included as a subalgebra. An example of such algebraic enlargement for the
Poincar\'{e} case was found in Refs.~\cite{Schrader, BCR}. The Maxwell
algebra presented there was used to describe the symmetries of quantum
fields in the Minkowski spacetime in a presence of the constant
electromagnetic field strength tensor. Its semisimple version appeared in
Refs.~\cite{Sorokas,Sorokas2} and represents the direct sum of the AdS and
Lorentz algebras.

Recently, the both mentioned examples, along with their supersymmetric
extensions, have been further extended by using generalized contractions
known as Lie algebra expansion methods \cite{HS, AIPV}. Together with a
later reformulation in terms of the abelian semigroups called the $S$-expansion \cite{Sexp}, these expansion methods proved to be a powerful tool
generating the new theories of gravity \cite{Zan, GRCS, GSRS, BDgrav,
DFIMRSV, CPRS2, SS} and supergravity \cite{deAzcarraga:2011pa, IRS1, FISV,
CRS}. Besides further studying the new supersymmetric schemes \cite{DGS, KL}, the subject finds also other applications. In Ref.~\cite{AKL} the
cosmological constant term in four dimensions arises from the Maxwell
algebra. The gauge fields related to the new generators might be useful in
inflation theories driven by the vector fields~\cite{AP} coupled to gravity
in a suitable way. Introduction of new fields and invariant tensors affects
the final form of the Lagrangians, which was particularly exploited in Refs.~\cite{Zan,GRCS} to establish a relation between General Relativity (GR) and
Chern-Simons (CS) gravity in odd dimensions. The same has been achieved for
even dimensions to relate Born-Infeld (BI) gravity with GR \cite{CPRS1}.
Other applications in the context of Bianchi algebras and a study of
properties of the $S$-expansion procedure with general semigroups have also
been analyzed in Refs.~\cite{CarocaNelson, AMNT}.

In this paper we introduce another family of Maxwell like algebras. Although
its existence could be understood in the $S$-expansion framework, we will
not write it explicitly from the very beginning. We will start by including
new generators $\left\{ Z_{ab},R_{a},...\right\} $ (with $a,b=1,\ldots,d$)
to the Lorentz and translational generators, adopting most of the conventions and
general setting presented in Ref.~\cite{SS}. We present a new scheme for
closing the enlarged algebras in a way different to already known Maxwell
families. By generalizing the change of basis from Ref.~\cite{Sorokas}, we
discover that the newly obtained algebras, denoted as $\mathfrak{D}_{m}$, can be seen as the direct sum of the AdS and the $\mathfrak{B}_{m-2}$ algebras
obtained by the expansion method \cite{GRCS}. In addition, $\mathfrak{D}_{m}$ algebras leads to $\mathfrak{B}_{m}$ under the In\"{o}n\"{u}-Wigner contraction. Finally, we explicitly incorporate these results within the $S$-expansion context and discuss the gravity actions in odd and even dimensions.

\section{Maxwell algebras}

The $S$-expansion procedure allows us to obtain two separate types of algebras, denoted in the literature as $\mathfrak{B}_{m}$ and $AdS\mathcal{L}_{m}$ \cite{GRCS, CPRS2, SS}. Both can be related with each other by the In\"{o}n\"{u}-Wigner contraction. Integer index $m>2$ labels different representatives, where standard generators of the Lorentz transformations $J_{ab}$ and translations $P_{a}$ become equipped with another set (or sets) of the new generators $Z_{ab}$ and $R_{a}$. Value $(m-1)$ might be used to indicate the total number of different generators $\{J_{ab},P_{a},Z_{ab}=Z_{ab}^{(1)},R_{a}=R_{a}^{(1)},Z_{ab}^{(2)},R_{a}^{(2)},\ldots\}$.

Our starting point is the Poincar\'{e} and AdS algebras, which can be
identified with $\mathfrak{B}_{3}$ and $AdS\mathcal{L}_{3}$, respectively.
Including the $Z_{ab}$ generator leads to $\mathfrak{B}_{4}$, which represents the Maxwell algebra \cite{Schrader, BCR}. It can be seen as a contraction of
the $AdS\mathcal{L}_{4}$ algebra \cite{Sorokas, SS}, in which the used notation resembles the fact that it combines the AdS with an additional Lorentz algebra. That algebra (along with its supersymmetric extension) originally appeared in \cite{Sorokas, Sorokas2} and was described as tensorial semi-simple enlargement of the Poincar\'{e} algebra. This was followed by a discussion of its deformations \cite{GKL}, and later reappeared in yet another form of
Maxwellian deformation of the AdS algebra~\cite{DKGS} with $
[P_{a},P_{b}]=(J_{ab}-Z_{ab})$. This last form reflects just the change of
basis with different decomposition of the generators $\{J_{ab},P_{a},Z_{ab}
\} $ forming the direct sum of $\mathfrak{so}(d-1,2)\oplus\mathfrak{so}(d-1,1)$
either out of AdS $\{Z_{ab},P_{a}\}\,\oplus$ Lorentz $\{(J-Z)_{ab}\}$ or AdS
$\{(J-Z)_{ab},P_{a}\}\,\oplus$ Lorentz $\{Z_{ab}\}$ generators. It would be
interesting to somehow relate this with the symmetry of fields, like was
done in Ref.~\cite{Schrader}, but now in the AdS spacetime with the constant electromagnetic field.

Name, originally used for $\mathfrak{B}_{4}$, was extended to describe
further generalizations for any index $m$, and also to take into account the
semi-simple Poincar\'{e} enlargement and its generalizations. Finally, they
all could be referred to as Maxwell type algebras \cite{CPRS2} or generalized
Maxwell algebras \cite{deAzcarraga:2012zv}. Since $AdS\mathcal{L}_{3}$
coincides with the AdS we find using label $AdS\mathcal{L}_{m}$ and the name
\textit{generalized AdS-Lorentz} from Ref.~\cite{SS} a little bit
misleading. Indeed, only in one case ($m=4$) we can talk about the direct sum of
AdS and Lorentz, while for $m>4$ the AdS algebra is no longer present as a
subalgebra. Therefore, throughout the paper we propose changing the label $
AdS\mathcal{L}_{m}$ into $\mathfrak{C}_{m}$, which will better fit the
scheme presented in this work.

In the case of $\mathfrak{B}_{5}$ and $\mathfrak{C}_{5}$ (formerly in \cite%
{SS} called as $AdS\mathcal{L}_{5}$) we can write their common part of the
commutation relations as
\begin{align}
\left[ P_{a},P_{b}\right] & =Z_{ab}\,,  \notag \\
\left[ J_{ab},P_{c}\right] & =\eta_{bc}P_{a}-\eta_{ac}P_{b}\,,  \notag \\
\left[ J_{ab,}J_{cd}\right] & =\eta_{bc}J_{ad}+\eta_{ad}J_{bc}-\eta
_{ac}J_{bd}-\eta_{bd}J_{ac}\,,  \notag \\
\left[ J_{ab,}Z_{cd}\right] & =\eta_{bc}Z_{ad}+\eta_{ad}Z_{bc}-\eta
_{ac}Z_{bd}-\eta_{bd}Z_{ac}\,,  \notag \\
\left[ Z_{ab},P_{c}\right] & =\eta_{bc}R_{a}-\eta_{ac}R_{b}\,,  \notag \\
\left[ J_{ab},R_{c}\right] & =\eta_{bc}R_{a}-\eta_{ac}R_{b}\,.
\label{B5C5shared}
\end{align}
When the algebra closes by satisfying
\begin{align}
\left[ R_{a},R_{b}\right] & =Z_{ab}\,,  \notag \\
\left[ Z_{ab},R_{c}\right] & =\eta_{bc}P_{a}-\eta_{ac}P_{b}\,,  \notag \\
\left[ Z_{ab,}Z_{cd}\right] & =\eta_{bc}J_{ad}+\eta_{ad}J_{bc}-\eta
_{ac}J_{bd}-\eta_{bd}J_{ac}\,,  \notag \\
\left[ R_{a},P_{b}\right] & =J_{ab}\,,  \label{C5}
\end{align}
we obtain $\mathfrak{C}_{5}$. After the following rescaling
\begin{equation}
P_{a}\rightarrow\mu P_{a}\,,\qquad Z_{ab}\rightarrow\mu^{2}Z_{ab}\,,\qquad
\text{and}\qquad R_{a}\rightarrow\mu^{3}R_{a}\,,  \label{Inonu}
\end{equation}
the In\"{o}n\"{u}-Wigner (IW) contraction \cite{IW} of the $\mathfrak{C}_{5}$
algebra in the limit of dimensionless parameter $\mu\rightarrow\infty$
shares exactly the form of common part (\ref{B5C5shared}), whereas the
remaining commutators become
\begin{align}
\left[ R_{a},R_{b}\right] & =0\,,  \notag \\
\left[ Z_{ab},R_{c}\right] & =0\,,  \notag \\
\left[ Z_{ab},Z_{cd}\right] & =0\,,  \notag \\
\left[ R_{a},P_{b}\right] & =0\,.  \label{B5}
\end{align}
It describes $\mathfrak{B}_{5}$, whose applications in Refs.~\cite{Zan,GRCS}
were already mentioned in the Introduction.

As we will see in the next section the separation on the two subsets of the
commutation relation is crucial to find a new algebra.

\section{Direct Maxwell algebras}

Intriguingly, there is one more way to close the subset of commutators
listed in (\ref{B5C5shared}), which is given by
\begin{align}
\left[ R_{a},R_{b}\right] & =Z_{ab}\,,  \notag \\
\left[ Z_{ab},R_{c}\right] & =\eta_{bc}R_{a}-\eta_{ac}R_{b}\,,  \notag \\
\left[ Z_{ab,}Z_{cd}\right] & =\eta_{bc}Z_{ad}+\eta_{ad}Z_{bc}-\eta
_{ac}Z_{bd}-\eta_{bd}Z_{ac}\,,  \notag \\
\left[ R_{a},P_{b}\right] & =Z_{ab}\,.  \label{D5}
\end{align}

The result, surprisingly, can be seen as a direct sum of two subalgebras. As
we will see, this example opens the whole new family of algebras where, in contrast to $\mathfrak{C}_{m>4}$, the AdS subalgebra is always present.
To show this, we introduce a generalization of the change of basis presented
in Ref.~\cite{Sorokas}, which now applies also to the ``translational'' generator.
With the group indices specified as $a,b=1,...,d$ we define two sets of
generators
\begin{equation}
L_{IJ}=\left\{
\begin{array}{c}
L_{ab}=Z_{ab}\,, \\
L_{a(D+1)}=R_{a}\,,
\end{array}
\qquad\text{and}\right. \qquad N_{IJ}=\left\{
\begin{array}{c}
N_{ab}=(J_{ab}-Z_{ab})\,, \\
N_{a}=(P_{a}-R_{a})\,,
\end{array}
\right.  \label{Soroka}
\end{equation}
satisfying the AdS
\begin{equation}
\left[ L_{IJ,}L_{KL}\right] =\eta_{JK}L_{IL}+\eta_{IL}L_{JK}-\eta
_{IK}L_{JL}-\eta_{JL}L_{IK}\,,
\end{equation}
and the Poincar\'{e} algebra
\begin{align}
\left[ N_{ab,}N_{cd}\right] & =\eta_{bc}N_{ad}+\eta_{ad}N_{bc}-\eta
_{ac}N_{bd}-\eta_{bd}N_{ac}\,,  \notag \\
\left[ N_{ab},N_{c}\right] & =\eta_{bc}N_{a}-\eta_{ac}N_{b},\qquad\left[
N_{a},N_{b}\right] =0\,.
\end{align}
It is straightforward to check that
\begin{equation}
\lbrack L_{IJ},N_{KL}]=0,
\end{equation}
therefore, they form the direct sum of $\mathfrak{so}(d-1,2)\oplus \mathfrak{iso}(d-1,1)$. From now on we will denote this algebra as $\mathfrak{D}_{5}$, where the used letter emphasizes the direct character of the found structure.

\bigskip Similarly, for one more generator, $Z_{ab}^{(2)}=\hat{Z}_{ab}$,
added to $\left\{ J_{ab},P_{a},Z_{ab},R_{a}\right\} $, the new algebra $
\mathfrak{D}_{6}$ will share with the $\mathfrak{B}_{6}$ and $\mathfrak{C}_{6}$ the same subset of commutators
\begin{align}
\left[ P_{a},P_{b}\right] & =Z_{ab}\,,  \notag \\
\left[ J_{ab},P_{c}\right] & =\eta_{bc}P_{a}-\eta_{ac}P_{b}\,,  \notag \\
\left[ J_{ab,}J_{cd}\right] & =\eta_{bc}J_{ad}+\eta_{ad}J_{bc}-\eta
_{ac}J_{bd}-\eta_{bd}J_{ac}\,,  \notag \\
\left[ J_{ab,}Z_{cd}\right] & =\eta_{bc}Z_{ad}+\eta_{ad}Z_{bc}-\eta
_{ac}Z_{bd}-\eta_{bd}Z_{ac}\,,  \notag \\
\left[ Z_{ab},P_{c}\right] & =\eta_{bc}R_{a}-\eta_{ac}R_{b}\,,  \notag \\
\left[ J_{ab},R_{c}\right] & =\eta_{bc}R_{a}-\eta_{ac}R_{b}\,,  \notag \\
\left[ Z_{ab,}Z_{cd}\right] & =\eta_{bc}\hat{Z}_{ad}+\eta_{ad}\hat {Z}_{bc}-\eta_{ac}\hat{Z}_{bd}-\eta_{bd}\hat{Z}_{ac}\,,  \notag \\
\left[ J_{ab,}\hat{Z}_{cd}\right] & =\eta_{bc}\hat{Z}_{ad}+\eta_{ad}
\hat{Z}_{bc}-\eta_{db}\hat{Z}_{bd}-\eta_{bd}\hat{Z}_{ac}\,,  \notag \\
\left[ P_{a},Z_{b}\right] & =\hat{Z}_{ab}\,.  \label{B6C6shared}
\end{align}
Additional rules reproducing $\mathfrak{C}_{6}$ are provided through
\begin{align}
\left[ \hat{Z}_{ab},P_{c}\right] & =\eta_{bc}P_{a}-\eta_{ac}P_{b}\,,
\notag \\
\left[ \hat{Z}_{ab},R_{c}\right] & =\eta_{bc}R_{a}-\eta_{ac}R_{b}\,,
\notag \\
\left[ Z_{ab,}\hat{Z}_{cd}\right] &
=\eta_{bc}Z_{ad}+\eta_{ad}Z_{bc}-\eta_{ac}Z_{bd}-\eta_{bd}Z_{ac}\,,  \notag
\\
\left[ \hat{Z}_{ab,}\hat{Z}_{cd}\right] & =\eta_{bc}\hat{Z}_{ad}+\eta_{ad}\hat{Z}_{bc}-\eta_{ac}\hat{Z}_{bd}-\eta_{bd}\hat{Z}_{ac}\,,  \notag \\
\left[ Z_{ab},R_{c}\right] & =\eta_{bc}P_{a}-\eta_{ac}P_{b}\,,  \notag \\
\left[ R_{a},R_{b}\right] & =Z_{ab}\,,  \label{C6}
\end{align}
while $\mathfrak{B}_{6}$ is achieved by the IW scaling (\ref{Inonu}) with
additional $\hat{Z}_{ab}\rightarrow\mu^{4}\hat{Z}_{ab}$, which forces the last part to commute. Finally, we can also find an alternative result of the
commutators satisfying the Jacobi identities,
\begin{align}
\left[ \hat{Z}_{ab},P_{c}\right] & =\eta_{bc}R_{a}-\eta_{ac}R_{b}\,,
\notag \\
\left[ \hat{Z}_{ab},R_{c}\right] & =\eta_{bc}R_{a}-\eta_{ac}R_{b}\,,
\notag \\
\left[ Z_{ab,}\hat{Z}_{cd}\right] & =\eta_{bc}\hat{Z}_{ad}+\eta_{ad}\hat{Z}_{bc}-\eta_{ac}\hat{Z}_{bd}-\eta_{bd}\hat{Z}_{ac}\,,  \notag \\
\left[ \hat{Z}_{ab,}\hat{Z}_{cd}\right] & =\eta_{bc}\hat{Z}_{ad}+\eta_{ad}\hat{Z}_{bc}-\eta_{ac}\hat{Z}_{bd}-\eta_{bd}\hat{Z}_{ac}\,,  \notag \\
\left[ Z_{ab},R_{c}\right] & =\eta_{bc}R_{a}-\eta_{ac}R_{b}\,,  \notag \\
\left[ R_{a},R_{b}\right] & =\hat{Z}_{ab}\,.  \label{D6}
\end{align}
This new $\mathfrak{D}_{6}$ algebra could be rewritten by the generalized
change of basis with
\begin{equation}
L_{IJ}=\{\hat{Z}_{ab},R_{a}\}
\end{equation}
forming the AdS, and
\begin{equation}
N_{IJ}=\{(J-\hat{Z})_{ab},(P-R)_{a},(Z-\hat{Z})_{ab}\}
\end{equation}
obeying nothing else than the $\mathfrak{B}_{4}$ algebra. Once again, we see that
\begin{equation}
\lbrack L_{IJ},N_{KL}]=0\,,
\end{equation}
therefore, we have obtained the direct sum $\mathfrak{D}_{6}=AdS\oplus
\mathfrak{B}_{4}$.

This can be easily generalized to the next examples. In fact, taking into
account that $\mathfrak{B}_{3}\equiv$\ Poincar\'{e} we conclude that we have
found the new class of algebras for$\ m>4$ (for at least four generators, $
J_{ab},P_{a},Z_{ab},R_{a},...$) being the direct sums
\begin{equation}
\mathfrak{D}_{m}=AdS\oplus\mathfrak{B}_{m-2}\,.
\end{equation}

As we remember, the family $\mathfrak{B}_{m}$ can be seen as the IW
contraction of the $\mathfrak{C}_{m}$ algebras. If we apply to $\mathfrak{D}_{m}$ the scaling from (\ref{Inonu}), now appended with new generators and
the further polynomial factors up to $\mu ^{m-2}$, then in a limit $\mu
\rightarrow \infty $ we obtain
\begin{equation}
\mathfrak{D}_{m}\rightarrow \mathfrak{B}_{m}\,.
\end{equation}
Thus, $\mathfrak{B}_{m}$ is also a limit in this case.\newline

\section{$S$-expansion and $\mathfrak{D}_{m}$ algebras}

Obtaining an explicit form of the commutators for arbitrary $m$ with the
procedure of the previous section is a straightforward but tedious task. However, the $S$-expansion method allows us to do it in a more compact way
using abelian semigroups (for further details see \cite{Sexp,SS}). To
reproduce the $\mathfrak{D}_{m}$ algebra we require that the elements of the relevant semigroup satisfy
\begin{equation}
\lambda _{\alpha }\lambda _{\beta }=\left\{
\begin{array}{l}
\lambda _{\alpha +\beta }\,,\qquad \qquad \qquad \qquad \qquad \qquad \quad
\,\,\,\text{for\quad }\alpha +\beta \leq m-2 \\
\lambda _{(\alpha +\beta -(m-1))\operatorname{modulo}2+(m-3)}\,,\qquad \qquad \text{for\quad }\alpha +\beta >m-2
\end{array}
\right.
\end{equation}
with the subset resonant decomposition $S_{0}\cup S_{1}$, where
\begin{align*}
S_{0}& =\left\{ 0_{S},\lambda _{0},\lambda _{2i}\right\} \qquad \text{ with } i=1,\ldots ,\left[ \frac{m-2}{2}\right] \,, \\
S_{1}& =\left\{ 0_{S},\lambda _{1},\lambda _{2j+1}\right\} \quad \text{ with
}j=1,\ldots ,\left[ \frac{m-3}{2}\right] \,.
\end{align*}
Note that the same commutation relations for the generators of $\mathfrak{B}
_{m},\mathfrak{C}_{m}$, and $\mathfrak{D}_{m}$ can be traced down to the
part with a standard multiplication rule $\lambda _{\alpha }\lambda _{\beta
}=\lambda _{\alpha +\beta }$.

The new algebra will be generated by the set of generators $\left\{
J_{ab},P_{a},Z_{ab}^{\left( i\right) },R_{a}^{\left( j\right) }\right\} $
related to the original AdS ones $\{\tilde{J}_{ab},\tilde{P}_{a}\}$ by
\begin{align*}
J_{ab}& =\lambda _{0}\times \tilde{J}_{ab},\qquad \,P_{a}=\lambda _{1}\times
\tilde{P}_{a}\,, \\
Z_{ab}^{\left( i\right) }& =\lambda _{2i}\times \tilde{J}_{ab},\qquad
R_{a}^{\left( j\right) }=\lambda _{2j+1}\times \tilde{P}_{a}\,,
\end{align*}
and satisfying
\begin{eqnarray}
\left[ J_{ab},J_{cd}\right]  &=&\eta _{bc}J_{ad}+\eta _{ad}J_{bc}-\eta
_{ac}J_{bd}-\eta _{bd}J_{ac}\,,  \notag \\
\left[ J_{ab},P_{c}\right]  &=&\eta _{bc}P_{a}-\eta _{ac}P_{b}\,,  \notag \\
\left[ P_{a},P_{b}\right]  &=&Z_{ab}^{\left( 1\right) }\,,  \notag \\
\left[
J_{ab},R_{c}^{\left( i\right) }\right] &=&\eta _{bc}R_{a}^{\left( i\right)
}-\eta _{ac}R_{b}^{\left( i\right) }\,,  \notag \\
\left[ J_{ab},Z_{cd}^{\left( i\right) }\right]  &=&\eta _{bc}Z_{ad}^{\left(
i\right) }+\eta _{ad}Z_{bc}^{\left( i\right) }-\eta _{ac}Z_{bd}^{\left(
i\right) }-\eta _{bd}Z_{ac}^{\left( i\right) }\,,  \notag \\
\left[ Z_{ab}^{\left( i\right) },Z_{cd}^{\left( j\right) }\right]  &=&\delta
_{2k}^{\left( 2i+2j-\left( m-1\right) \right) \operatorname{mod}2+(m-3)}\left( \eta
_{bc}Z_{ad}^{\left( k\right) }+\eta _{ad}Z_{bc}^{\left( k\right) }-\eta
_{ac}Z_{bd}^{\left( k\right) }-\eta _{bd}Z_{ac}^{\left( k\right) }\,\right)
\text{\thinspace },  \notag \\
\left[ Z_{ab}^{\left( i\right) },R_{c}^{\left( j\right) }\right]  &=&\delta
_{2k+1}^{\left( 2i+2j-\left( m-2\right) \right) \operatorname{mod}2+(m-3)}\left(
\eta _{bc}R_{a}^{\left( k\right) }-\eta _{ac}R_{b}^{\left( k\right) }\right)
\,,  \notag \\
\left[ Z_{ab}^{\left( i\right) },P_{c}\right]  &=&\delta _{2k+1}^{\left(
2i-\left( m-2\right) \right) \operatorname{mod}2+(m-3)}\left( \eta
_{bc}R_{a}^{\left( k\right) }-\eta _{ac}R_{b}^{\left( k\right) }\right) \,,
\notag \\
\left[ R_{a}^{\left( i\right) },R_{b}^{\left( j\right) }\right]  &=&\delta
_{2k}^{\left( 2i+2j-\left( m-3\right) \right) \operatorname{mod}2+(m-3)}Z_{ab}^{\left( k\right) }\,,\text{ }  \notag \\
\left[ P_{a},R_{b}^{\left( i\right) }\right]  &=&\delta _{2k}^{\left(
2i-\left( m-3\right) \right) \operatorname{mod}2+(m-3)}Z_{ab}^{\left( k\right) }\,.
\end{eqnarray}
In addition, the direct sum of $\mathfrak{D}_{m}=\mathfrak{so}\left(
d-1,2\right) \oplus \mathfrak{B}_{m-2}$ will be explicitly composed from
\begin{equation}
L_{IJ}=\left\{ Z_{ab}^{\left(\left[ \left( m-2\right) /2\right]\right) },R_{a}^{\left(\left[
\left( m-3\right) /2\right]\right) }\right\} \,,  \label{L_gen}
\end{equation}
forming the AdS algebra, and
\begin{equation}
N_{IJ}=\left\{ \left( J-Z^{\left(\left[ \left( m-2\right) /2\right]\right) }\right)
_{ab},\left( P-R^{\left(\left[ \left( m-3\right) /2\right]\right) }\right) _{a},\left(
Z^{\left( i\right) }-Z^{\left(\left[ \left( m-2\right) /2\right]\right) }\right)
_{ab},\left( R^{\left( j\right) }-R^{\left(\left[ \left( m-3\right) /2\right]\right)
}\right) _{a}\right\} \,,  \label{N_gen}
\end{equation}
satisfying the $\mathfrak{B}_{m-2}$ algebra. This generalized change of basis, originating from result of Ref.~\cite{Sorokas}, remarkably reveals non-obvious structure of the algebra, showing that
the new way of closing Maxwell like algebras is nothing more than a direct
sum of already known algebras, $AdS$ and $\mathfrak{B}$.

Finally, we conclude this section with the summary
\begin{equation*}
\begin{tabular}{|c|l|c|c|c|}
\hline
$m$ & Generators & ~\qquad Type $\mathfrak{B}_{m}\qquad $ & Type $\mathfrak{C}_{m}$ & Type $\mathfrak{D}_{m}$ \\ \hline
3 & $J_{ab},P_{a}$ & $\mathfrak{B}_{3}=$Poincar\'{e} & $\mathfrak{C}_{3}=$AdS
& - \\ \hline
4 & $J_{ab},P_{a},Z_{ab}$ & $\mathfrak{B}_{4}=$Maxwell & $\mathfrak{C}_{4}=$
AdS$\oplus $Lorentz & - \\ \hline
5 & $J_{ab},P_{a},Z_{ab},R_{a}$ & $\mathfrak{B}_{5}$ & $\mathfrak{C}_{5}$ & $
\mathfrak{D}_{5}=$AdS$\oplus $Poincar\'{e} \\ \hline
6 & $J_{ab},P_{a},Z_{ab},R_{a},\tilde{Z}_{ab}$ & $\mathfrak{B}_{6}$ & $
\mathfrak{C}_{6}$ & $\mathfrak{D}_{6}=$AdS$\oplus $Maxwell \\ \hline
\ldots  & \ldots  & \ldots  & \ldots  & \ldots  \\ \hline
$m$ & $J_{ab},P_{a},Z_{ab}^{(i)},R_{a}^{(j)}$ & $\mathfrak{B}_{m}$ & $
\mathfrak{C}_{m}$ & $\mathfrak{D}_{m}=$AdS$\oplus \mathfrak{B}_{m-2}$ \\
\hline
\end{tabular}
\ \ \ \ \ \
\end{equation*}
where $Z_{ab}^{(i)}$ and $R_{a}^{(j)}$ represent the set of new generators
starting from $i,j=1$ and then preserving an order $Z^{(1)},R^{(1)},Z^{(2)},
\ldots $ to fill out a total number of $(m-1)$ generators.

\section{Gravity theories based on $\mathfrak{D}_{m}$ algebras}

One can use the $\mathfrak{D}_{m}$ algebras spanned by generators $\tilde{T}_{M}=\left\{ J_{ab},P_{a},Z_{ab}^{\left( i\right) },R_{a}^{\left( j\right)
}\right\} $ to construct the gravity theories. Depending on the dimensions,
it is possible to build the Chern-Simons or Born-Infeld actions based on the
one-form gauge connection
\begin{equation}
A=\tilde{A}^{M}\tilde{T}_{M}=\frac{1}{2}\tilde{\omega}^{ab}J_{ab}+\frac{1}{\ell }\tilde{e}^{a}P_{a}+\frac{1}{2}\tilde{k}^{ab\left( i\right)
}Z_{ab}^{\left( i\right) }+\frac{1}{\ell }\tilde{h}^{a\left( j\right)
}R_{a}^{\left( j\right) }\,,
\end{equation}
with $\ell $ being a length parameter introduced in order to have a
dimensionless one-form connection. The Lagrangians constructed with this
connection and with the corresponding invariant tensors, provided by means
of the $S$-expansion, will have the form
\begin{equation}
\mathcal{L}_{\mathfrak{D}_{m}}\left[ \tilde{\omega},\tilde{e},\tilde{k},
\tilde{h}\right] =\mathcal{L}\left[ \tilde{\omega},\tilde{e}\right] +
\mathcal{L}_{\text{int}}\left[ \tilde{\omega},\tilde{e},\tilde{k},\tilde{h}
\right] \,,
\end{equation}
where the last term, in general, contains non-trivial interactions between
gravity sector and extra fields associated with the new generators $
Z_{ab}^{\left( i\right) }$ and $R_{a}^{\left( j\right) }$.

However, as we have seen in the last section, a particular basis $T_{M}=\left\{
L_{IJ},N_{IJ}\right\} $ given by Eqs.~(\ref{L_gen}) and (\ref{N_gen}) shows
that the new algebras are actually direct sums, $\mathfrak{D}_{m}=AdS\oplus
\mathfrak{B}_{m-2}$, of algebras whose invariant tensors are already known
in the literature. Then, $A=\tilde{A}^{M}\tilde{T}_{M}=A^{M}T_{M}$
induces a redefinition of the fields making the interaction terms no
longer present. Indeed, the gauge connection in the direct basis is given by
\begin{equation}
A=A^{M}T_{M}=\frac{1}{2}\varpi ^{IJ}L_{IJ}+\frac{1}{2}\omega ^{IJ}N_{IJ}\,,
\end{equation}
where the field content explicitly corresponds to
\begin{align}
L_{IJ}\quad \rightarrow \quad \varpi ^{IJ}& =\{\varpi ^{ab},\frac{1}{\ell }
\bar{e}^{a}\}\,,  \label{field_content_L} \\
N_{IJ}\quad \rightarrow \quad \omega ^{IJ}& =\{\omega ^{ab},\frac{1}{\ell }
e^{a},k^{ab(i)},\frac{1}{\ell }h^{a(j)}\}\,.  \label{field_content_N}
\end{align}
Since our algebra is described as a direct sum, the related Lagrangian will
be written as the combination
\begin{equation}
\mathcal{L}_{\mathfrak{D}_{m}}\left[ \omega ,e,\varpi ,\bar{e},k,h\right] =
\mathcal{L}_{AdS}\left[ \varpi ,\bar{e}\right] +\mathcal{L}_{\mathfrak{B}_{m-2}}\left[ \omega ,e,k,h\right] \,,
\end{equation}
where each independent part will come with the separate set of constants
related with different components of the invariant tensors. In odd
dimensions the theory will be provided by the Chern-Simons form invariant
under the whole $\mathfrak{D}_{m}$ symmetry, while in even dimensions the
theory will be only invariant under the local Lorentz like subalgebra
reproducing Born-Infeld gravity.

Although it seems natural to identify the true vielbein with the component
associated with the AdS like $L_{IJ}$ generator satisfying standard
commutation relations, that choice is not the best one. Since the IW
contraction of $\mathfrak{D}_{m}$ recovers the $\mathfrak{B}_{m}$
algebra, we should identify the true vielbein with the field component
associated with the corresponding part of the $N_{IJ}$ generator, which
after contraction reproduces $P_{a}$ generator. The same argument holds for
the spin connection.

\subsection{$\mathfrak{D}_{m}$-invariant Chern-Simons gravity}

The Chern-Simons Lagrangian \cite{Cham1} in $d=(2n+1)$ spacetime is defined
as
\begin{equation}
\mathcal{L}_{CS}^{2n+1}[A]=\kappa (n+1)\int_{0}^{1}\delta t\left\langle
A\left( tdA+t^{2}A^{2}\right) ^{n}\right\rangle \,.
\end{equation}
Here, $\left\langle \ldots \right\rangle $ denotes an invariant tensor,
whose AdS components are given by
\begin{equation}
\left\langle J_{a_{1}a_{2}}\cdots
J_{a_{2n-1}a_{2n}}P_{a_{2n+1}}\right\rangle =\frac{2^{n}}{n+1}\epsilon
_{a_{1}a_{2}\cdots a_{2n+1}}\,,
\end{equation}
while according to Refs.~\cite{GRCS, CPRS3}, the non-vanishing components of an invariant tensor for the $\mathfrak{B}_{m}$ generators $
N_{IJ}=\{J_{ab,(i)},P_{a,(j)}\}$ are given by
\begin{equation}
\left\langle J_{a_{1}a_{2},(i_{1}) }\cdots J_{
a_{2n-1}a_{2n},(i_{\left( m-1\right) /2}) }P_{ a_{2n+1},(i_{\left(
m+1\right) /2}) }\right\rangle =\frac{2^{n}}{n+1}\alpha _{j}\delta
_{i_{1}+...+i_{\left( m+1\right) /2}}^{j}\epsilon _{a_{1}a_{2}\cdots
a_{2n+1}}\,.
\end{equation}
Thus, the AdS like counterpart of the full action yields the standard form
\begin{equation}
\mathcal{L}_{AdS}^{\left( 2n+1\right) }=\epsilon _{a_{1}\cdots
a_{2n+1}}\sum_{k=0}^{n}\frac{2}{\ell^{2\left( n-k\right) +1}}\left(
\begin{array}{c}
n \\
k
\end{array}
\right) \frac{1}{2\left( n-k\right) +1}\mathfrak{R}^{a_{1}a_{2}}(\varpi
)\cdots \mathfrak{R}^{a_{2k-1}a_{2k}}(\varpi )\bar{e}^{a_{2k+1}}\cdots \bar{e}^{a_{2n+1}}\,,
\end{equation}
whereas a construction of the $\mathfrak{B}_{m-2}$ CS theory can be found in Refs.~\cite{Zan, GRCS, CPRS2, CPRS3}.

It is worth analyzing in more detail the $\mathfrak{B}_{m-2}$ sector where, as pointed out in a previous section, the gravitational terms are present. Let us note that the Einstein-Hilbert action can be derived from the $\mathfrak{B}_{m}$ CS action when the matter fields are switched off. However, as was mentioned in Ref.~\cite{Zan}, the field equations do not reproduce the Einstein equations. Indeed, the variation of the action with respect to the vielbein, the spin connection and the matter fields leads to a much stronger restriction on the geometry, where the metric must satisfy simultaneously the EH, as well as all the Lovelock equations. Such restriction admits, besides the trivial flat spacetime, the pp-wave solutions.

Nevertheless, there is a particular configuration within the semigroup expansion formalism, which permits the recovery of the GR dynamics, as well as the Einstein action from a CS gravity action using the $\mathfrak{B}_{m}$ algebras. It comes from the fact that the invariant tensor constructed using the $S$-expansion procedure allows the introduction of a coupling constant $\ell$. As was shown in Ref.~\cite{GRCS}, considering $\ell =0$ and performing a matter-free configuration limit gives not only the GR action but
also the appropriate Einstein field equations. Now, by vanishing of the extra fields, we can also get rid of the contribution from an additional CS piece corresponding to AdS part of the $\mathfrak{D}_m$ algebra. Note that such setup allowing derivation of standard General Relativity can be applied not only in odd but also in even dimensions considering a Born-Infeld like gravity theory \cite{CPRS2, CPRS1}.

The original basis $\tilde{T}
_{M} = \left\{ J_{ab}, P_{a}, Z_{ab}^{(i)}, R_{a}^{(i)} \right\} $ would give a less straightforward picture. Indeed, the CS action based on the $\tilde{T}_{M}$ generators would lead to the more convoluted form of constraints on the geometry being much harder to study. Luckily using the new basis $T_{M}=\left\{L_{IJ},\ N_{IJ}\right\}$ of the $\mathfrak{D}_{m}$ algebras enable us to get a better insight using building blocks already known in the literature.

\subsection{$\mathbb{L}_{m}$-invariant Born-Infeld gravity}

In even dimensions $d=(2n+2)$ we focus on the Lorentz like algebra $\mathbb{L}_{m}=\mathfrak{so}\left( d-1,1\right) \oplus\mathfrak{L}_{m-2}$ being a
subalgebra of the $\mathfrak{D}_{m}$. In particular, the generators $L_{ab}$
will satisfy the Lorentz subalgebra of the $AdS$, while $N_{ab\left( i\right) } $ will satisfy the Lorentz like
subalgebra $\mathfrak{L}_{m-2}\subset\mathfrak{B}_{m-2}$.

The usual Born-Infeld gravity Lagrangian \cite{TZ} can be constructed from
the Lorentz components of the AdS curvature,
\begin{equation}
\mathcal{L}_{Lorentz}^{\left( 2n+2\right) }=\sum_{p=0}^{n}\left(
\begin{array}{c}
n+1 \\
p
\end{array}
\right) \ell^{2\left( p-n\right) -1}\epsilon_{a_{1}\dots a_{2n+2}}\mathfrak{R}
^{a_{1}a_{2}}(\varpi)\cdots\mathfrak{R}^{a_{2p-1}a_{2p}}(\varpi)\ \bar{e}
^{a_{2p+1}}\cdots\bar{e}^{a_{2n+2}}\,.
\end{equation}
Here, $\bar{e}^{a}$ corresponds to the vielbein type field and $\mathfrak{R}
^{ab}=d\varpi^{ab}+\varpi_{\text{ }c}^{a}\varpi^{cb}$ to the Riemann
curvature in the first-order formalism for a connection from Eq.~(\ref{field_content_L}). Although the above Lagrangian is constructed out of the AdS like curvatures it is invariant only under a Lorentz subgroup, thus, it
does not correspond to a topological invariant.

From Ref.~\cite{CPRS1}, it follows that the BI Lagrangian invariant under
the subalgebra $\mathfrak{L}_{m-2}$ of $\mathfrak{B}_{m-2}$ can be expressed
as
\begin{align}
\mathcal{L}_{\mathfrak{L}_{m-2}}^{\left( 2n+2\right) } &
=\sum_{k=1}^{n}\ell^{2\left( k-n\right) -1}\frac{1}{2n+2}\left(
\begin{array}{c}
n+1 \\
k
\end{array}
\right)
\alpha_{j}\delta_{i_{1}+\ldots+i_{n+1}}^{j}\delta_{p_{1}+q_{1}}^{i_{k+1}}
\cdots\delta_{p_{n+1-k}+q_{n+1-k}}^{i_{n+1}}  \notag \\
& \varepsilon_{a_{1}\cdots a_{2n+2}}R^{ a_{1}a_{2},(i_{1})
}\cdots R^{ a_{2k-1}a_{2k},(i_{k}) }e^{
a_{2k+1},(p_{1}) }e^{ a_{2k+2},(q_{1}) }\cdots e^{
a_{2n+1},(p_{n+1-k}) }e^{ a_{2n+2},(q_{n+1-k}) }\,.
\label{EBI}
\end{align}

After combining these two parts, we find that the $\left( 2n+2\right) $
-dimensional Born-Infeld Lagrangian invariant under a Lorentz like
subalgebra of the $\mathfrak{D}_{m}$ algebra can be decomposed as the sum
\begin{equation}
\mathcal{L}_{\mathbb{L}_{m}}^{\left( 2n+2\right) }=\mathcal{L}_{Lorentz}^{\left( 2n+2\right) }+\mathcal{L}_{\mathfrak{L}_{m-2}}^{\left(
2n+2\right) }\,.
\end{equation}

\section{Conclusion}

Although the new generators and introduced modifications still require
physical interpretation, the Maxwell algebras of all types with their
supersymmetric extensions have found interesting applications in
(super)gravity. They were used to study interrelations between different
theories with the main focus concerning non-direct sum structures, which
admit non-trivial modifications of the dynamics. Curiously, the new family
of Maxwell algebras $\mathfrak{D}_{m}$ derived here, under an appropriate
change of basis, turned out to be a direct sum of the $AdS$ and $\mathfrak{B}
_{m-2}$ algebras. It still would be interesting to analyze our new algebra
in the same context as was done for semi-simple Maxwell algebra $\mathfrak{C}_{4}=$AdS$\oplus$Lorentz, at least for the cases where to the AdS we add the Poincar\'{e} or the Maxwell algebra.

A minimal supersymmetric version of the $\mathfrak{D}_{m}$ algebra could
avoid the problematic fermionic anticommutator present in non-standard
supersymmetric extension of Maxwell algebras \cite{Sorokas2, Lukierski, CFRS}. On the other hand, an alternative family of minimal Maxwell superalgebras
would lead to the direct sum of the minimal Maxwell superalgebras introduced
in Refs.~\cite{CR1, CR2, Bogokalu} and the super-AdS ones, thus enlarging
the D'Auria-Fr\'{e} and the Green \cite{AF, Green} superalgebras.

Additionally, we have provided the abelian semigroup expansion procedure
leading to $\mathfrak{D}_{m}$, which allowed us for a straightforward
construction of the gravity actions. Due to the direct structure of the
algebra we have shown that the interaction terms can be always removed by a
redefinition of the fields, which results in somehow trivial gravity models.
However, under some limits it still can lead to the interesting results.

The obtained family of algebras needs an interpretation not only for description
of the new symmetries and their consequences. Existence of the $\mathfrak{D}
_{m}$ algebra, which under the In\"{o}n\"{u}-Wigner contraction leads to $
\mathfrak{B}_{m}$, exactly as it happened for the $\mathfrak{C}_{m}$
algebra, makes us consider the possibility of finding other semigroups
useful for gravity models.

\section{Acknowledgment}

We thank J. Lukierski, O. Mi\v{s}kovi\'{c} and P. Salgado for valuable
comments and discussion. This work was supported by the Chilean FONDECYT
Projects No. 3140267 (RD) and 3130445 (NM), and also funded by the Newton-Picarte CONICYT Grant No. DPI20140053 (PKC and EKR).

\end{document}